\documentclass[journal]{IEEEtran}
\IEEEoverridecommandlockouts
\usepackage{cite}

\usepackage{amsmath,amssymb,amsfonts,mathtools,bm}
\usepackage{amsthm}
\usepackage{algorithm}
\usepackage{algorithmic}
\usepackage{graphicx}
\usepackage{textcomp}
\usepackage{xcolor,float}
\usepackage{tabularx}
\usepackage{textcomp}
\usepackage{diagbox}
\usepackage{svg}
\usepackage{booktabs}

\allowdisplaybreaks
\renewcommand{\baselinestretch}{1}

\begin{document}

    \title{{On-Demand Resource Management for 6G Wireless Networks Using Knowledge-Assisted Dynamic Neural Networks}}
\author{
\IEEEauthorblockN{
Longfei Ma\IEEEauthorrefmark{1},
Nan Cheng\IEEEauthorrefmark{1},
Xiucheng Wang\IEEEauthorrefmark{1}, 
Ruijin Sun\IEEEauthorrefmark{1} and
Ning Lu\IEEEauthorrefmark{2}\\
}
\IEEEauthorblockA{
\IEEEauthorrefmark{1}School of Telecommunications Engineering,
Xidian University, Xi'an, China\\
\IEEEauthorrefmark{2}Department of Electrical and Computer Engineering, Queen’s University, Kingston K7L 3N6, Ontario, Canada\\
Email: lfma@stu.xidian.edu.cn, nancheng@xidian.edu.cn, xcwang\_1@stu.xidian.edu.cn, sunruijin@xidian.edu.cn, ning.lu@queensu.ca}}

    \maketitle

\IEEEdisplaynontitleabstractindextext

\IEEEpeerreviewmaketitle

\begin{abstract}
On-demand service provisioning is a critical yet challenging issue in 6G wireless communication networks, since emerging services have significantly diverse requirements and the network resources become increasingly heterogeneous and dynamic. In this paper, we study the on-demand wireless resource orchestration problem with the focus on the computing delay in orchestration decision-making process. Specifically, we take the decision-making delay into the optimization problem. Then, a dynamic neural network (DyNN)-based method is proposed, where the model complexity can be adjusted according to the service requirements. We further build a knowledge base representing the relationship among the service requirements, available computing resources, and the resource allocation performance. By exploiting the knowledge, the width of DyNN can be selected in a timely manner, further improving the performance of orchestration.  Simulation results show that the proposed scheme significantly outperforms the traditional static neural network, and also shows sufficient flexibility in on-demand service provisioning. 
\end{abstract}

\begin{IEEEkeywords}
6G, on-demand service, dynamic neural network, knowledge, resource orchestration

\end{IEEEkeywords}

\section{Introduction}

Over the recent years, the fifth generation (5G) wireless communication networks have been commercialized and deployed worldwide. Different from the previous wireless communication systems which merely focus on broadband communications, 5G networks support three major communication scenarios, i.e., enhanced mobile broadband (eMBB), ultra-reliable and low latency communications (URLLC), and massive machine type communications (mMTC). By defining these scenarios with individually focusing performance indicators, 5G networks enable a distinctive capability to provide improved network services not only to mobile users, but also to vertical fields such as industry, education, and automobile \cite{7414384,8304385}. By means of network slicing, the network resources can be allocated to individual services or users in an isolated way, which guarantees the service performance level. Although 5G networks can achieve much improved network performance and guarantee service-level performance, there are still some limitations. For example, mainly relying on terrestrial base stations which are densely deployed in urban areas but not in rural or remote areas, the 5G networks can hardly provide efficient services in some remote and harsh areas. Therefore, the applications in these areas are significantly limited, such as IoT monitoring, autonomous vehicles, telesurgery, etc \cite{8672604}. 

With the vigorous development of emerging technologies such as space-air-ground integrated networks \cite{8672604,7981533}, terahertz \cite{8732419} and intelligent reflecting surface \cite{8910627}, as well as the deep integration of artificial intelligence and communication technologies, it provides a broad prospect for the research of the sixth generation (6G) wireless communication networks. One of the significant challenges of 6G network is to offer “on-demand” services, i.e., on-demand service provisioning. On one hand, due to the wide coverage and the application of software-defined technologies and AI technologies, a soaring number of novel services with diverse requirements will emerge. These services and requirements should be efficiently and cost-effectively handled and satisfied. On the other hand, the high network dynamics, network architecture heterogeneity, and network resource complexity will jointly result in unprecedented difficulty in network resource orchestration. To address these issues, 6G networks are envisioned to comprehensively exploit new air interface, novel network architecture, and most importantly, advanced AI technologies to further improve network performance and offer on-demand services. 

In this paper, we focus on a paramount but not yet widely studied issue in providing on-demand services, which is the decision-making process in network resource orchestration. Apparently, a fast and high-optimality decision is preferred to achieve higher network performance. However, in most cases, the speed (time required for the decision-making process) and the optimality performance (gap between the decision and the global optimal decision) are tradeoff metrics, and it would be very promising if the tradeoff can be adjusted according to the specific scenario and service requirements. The common way is to formulate the decision-making problem as an optimization problem, and then solve the problem by iteration-based methods, such as Lagrangian method, successive convex approximation (SCA) and alternating direction method of multipliers (ADMM). However, using the iteration-based methods may lead to large computational complexity, and therefore confronted with the larger network scale and higher network dynamics of 6G. Furthermore, such methods may result in excessive delay to converge, or even impossible to solve the problem. According to the theory in \cite{universal}, neural network can approach any continuous function infinitely. In addition, compared with the traditional iteration-based algorithms, the inference computation of neural networks is lower and hence the decision takes less time, due to the fact that a multi-layer neural network only needs a fixed number of matrix multiplications to get the inference results \cite{goodfellow2016deep}. Therefore, by reasonably setting and training the network structure, we can also use neural network to solve the above network resource orchestration problem efficiently, e.g., using the deep neural network for complex resource allocation \cite{liu2018deep,8680025}, or deep reinforcement learning for dynamic network decision-making\cite{9222519}. However, in the era of 6G, the diverse scenario and service requirements may lead to different and changing demands for the resource orchestration decision-making. For example, a base station serving self-driving cars will require much lower decision-making delay than serving IoT monitoring. In addition, the amount of available computing resources on the base station will make the issue more challenging. Most neural network models do not offer such flexibility that the decision speed and optimality can be adjusted according to the service requirement and available computing resources.

Recently, some scholars have proposed dynamic neural network (DyNN), which selectively executes different parts of the neural network according to different inputs, solving the problem of balancing between model inference delay and calculation accuracy \cite{moe}. Such model flexibility renders DyNN very promising in 6G on-demand service provisioning. In this paper, we propose a DyNN-based resource orchestration scheme to solve the on-demand decision-making problem for resource orchestration in 6G wireless networks. We formulate a multi-service bandwidth allocation problem, where each service has its own preference on service delay. The decision network (i.e., the DyNN model) can dynamically adjust the network width according to the characteristics of tasks and the different computing power of users and give the optimal solution under the comprehensive consideration of inference delay and transmission delay. Specifically, we divide the bandwidth allocation problem into two stages. First, the optimal width of the decision network is determined according to the task characteristics and user computing power by exploiting a constructed knowledge base. Then, the optimal bandwidth allocation scheme is obtained by inputting the task characteristics to the optimal decision network.

Our main contributions are three-fold.
(1) We use a dynamic neural network whose network width can vary with task characteristics to realize on-demand resource allocation.
(2) The influence of device computing capability on the performance of allocation algorithm is studied.
(3) A knowledge base of network width, task characteristics and equipment performance is established, and the knowledge-assisted network width adjustment is realized.

\section{System Model}
\subsection{Communication and Computing Model}
We consider a single-cell multi-user downlink communication scenario. In this case, each geographically distributed user generates a downlink transmission task in each communication frame.
Specifically, we consider $N$ single-antenna users, where user $i$ is denoted by $u_{i}$ and the distance between $u_{i}$ and the cell base station is denoted by $d_{i}, i \in {1,2,...N}$.  
At the beginning of each frame $t$, each user $u_{i}$ requests a downlink transmission task, denoted by $o_{i}$. Consider that a task $o_{i}$ is described by two parameters, i.e., the task size $s_{i}$ and task importance weight $w_{i}$. Note that the  task importance weight $w_{i}$ represents the preference of user $i$ on the service delay, with a higher value of $w_{i}$ representing that the user requires a lower service delay. The available computing power of the base station used for resource orchestration is denoted by $f$. Note that $f$ is changing in different frames due to multiple reasons, such as energy constraints and computing power used by other tasks. We consider that the system uses frequency division multiple access (FDMA), where the total bandwidth is denoted by $b_{max}$. In each frame, the base station allocates bandwidth to each users. We define the weighted service transmission delay $T_{tra,i}$ of user $u_{i}$ during a frame as
\begin{equation*}
    T_{tra,i} =  w_i \frac{s_i}{b_{i} \log_2 \left(1+\frac{P  (\frac{d_0}{d_{i}})^2 g_0}{\sigma^2} \right)}, \tag{1}\label{Ttra}
\end{equation*}
where $b_{i}$ is bandwidth to transmit task to $u_i$, $P$ and $\sigma^2$ are respectively transmitting power and noise power, $g_0$ and $d_0$ are respectively the unit channel gain and distance between users and base station. Then, the total weighted service transmission time $T_{tra}$ can be calculated as $\sum_{i=1}^{N} T_{t r a, i}$

In this paper, a neural network (NN) is exploited for determining the resource orchestration scheme. A $L$-layers NN can be denoted by $G(x) = g^L(\cdots (g^2(g^1(x))))$, where $g^j(x) = h^j(\mathbf{A}x +\mathbf{B}),\forall j \in \{1,2,\cdots,L\}$, $h^j(\cdot)$ is a non-linear function, $\mathbf{A}$ is a parameter matrix, and $\mathbf{B}$ is parameter vector, whose size is determined by the nodes of $j$-1-th layer and $j$-th layer. According to \cite{goodfellow2016deep}, the computing complexity of a $m$ nodes layer is $\mathcal{O}(m)$, which is proportional to the parameter quantity. Therefore, considering the importance of the task $o_{i}$, the weighted inference computing delay $T_{com,i}$ for a $L$-layers NN can be calculated by
\begin{equation*}
    T_{com,i} = w_{i}\sum_{j=1}^{L-1} \alpha \frac{\left\|\theta_{j}\right\|\left\|\theta_{j+1}\right\|}{f},\tag{2}\label{Tcom}
\end{equation*}
where $\left\|\theta_{j}\right\|$ is the parameter quantity of $j$-th layer and $f$ is CPU frequency, and $\alpha$ is computing efficiency factor which is determined by NN implementing method and CPU architecture. Then, the total weighted inference delay $T_{com}$ can be calculated as $\sum_{i=1}^{N} T_{com,i}$

\subsection{Problem Formulation}
The architecture of NN (width or depth) can significantly impact the model fitting ability and thus the inference accuracy. Generally, a wider/deeper NN can achieve a relatively higher inference accuracy, and thus the resulting resource allocation has higher performance, e.g., lower transmission delay. On the other hand, according to (2), the computing delay will also increase with increasing NN width or depth. Therefore, the total service delay, including the computing delay $T_{com}$ and the transmission delay $T_{tra}$, might be even higher with a wider/deeper NN especially when the task size is small, since the computing delay becomes the dominant part. In this paper, we propose a DyNN-based resource allocation, where the width/depth of NN can be adjusted to balance the service transmission delay and computing delay. We formulate the bandwidth allocation scheme as an optimization problem in order to minimize the total weighted service delay within a cell as follows.
\begin{align*}
    & min_{\mathbf{b},\mathbf{\theta}}\quad T_{tra} + T_{com} \tag{3} \\
    &s.t.\quad\quad\quad \sum_{i} b_{i}\leq b_{max}, \tag{3a} \\
    &\quad\quad\quad\quad\quad \left\|\theta_{j}\right\| > 0\quad \forall j \in \{1,2,\cdots,L\}, \tag{3b}\\
    &\quad\quad\quad\quad\quad \mathbf{b} = G(\mathbf{w},\mathbf{s},\mathbf{d};\mathbf{\theta}), \tag{3c}
\end{align*}
where $\mathbf{b}$ is the bandwidth allocation scheme, $\mathbf{w}$, $\mathbf{s}$ and $\mathbf{d}$ are respectively the vectors of task importance, task size and transmission distance, $G(\cdot)$ and $\theta$ are the mapping function and parameter of NN respectively.

\section{Knowledge Assisted Dynamic Neural Network Method}
\subsection{Dynamic Neural Network}
Taking a close look at the optimization problem (3), we can find that the total service delay is affected by several factors, e.g., the service requirements, the task size, and the computing capability. For instance, when the task size is small, or the overall service delay requirement is very strict, the computing delay becomes the dominant delay component, and thus a NN with lower width is preferred to reduce the computing delay (also the total delay), and vice versa. In addition, when the available computing power of the base station is small, a smaller NN is also preferred to achieve a lower computing delay. However, although adjusting the NN size is important for minimizing the service delay, it is usually difficult as the NN size (e.g., number of layers or layer size) is a hyperparameter.

A naive idea is to train multiple NNs with different widths independently, and select a suitable NN to allocate resource according to the delay requirement or task size. However, training multiple NNs will not only increase the training time and cost, but also lead to degrading training performance and inference accuracy, since the sparse training samples obtained from the physical interactions with the wireless environments will be insufficient for training many different NNs. Moreover, a light-weight node such as an IoT node is usually equipped with weak computing and storage capabilities, and caching many NNs may overwhelm the storage of a node.

To efficiently address the issue, we proposed a variable-width DyNN based resource allocation scheme, which can adjust the hidden layers width according to the task size and the user's computing capability. When the task size is small or the user's computing capability is insufficient, a hidden layer with a smaller width is used to reduce the inference(computing) delay. When the task is large, a hidden layer with a large width is used to improve the inference accuracy and reduce the transmission delay, thereby reducing the overall service delay.

\begin{figure}
  \centering
  \includegraphics[width=0.9\columnwidth]{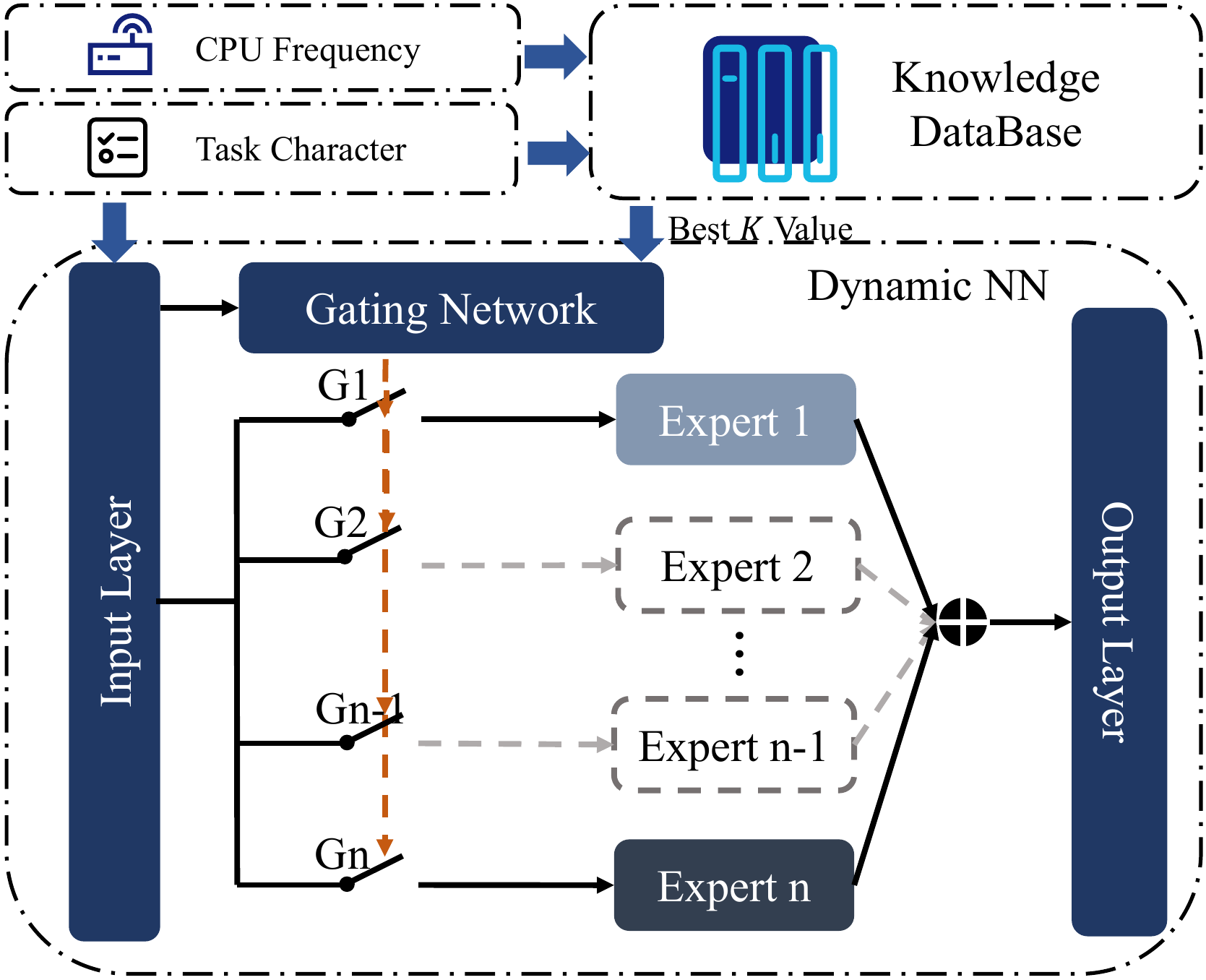}
   \vspace{-5pt}
  \caption {Overview of the proposed DyNN-based method.}
  \label{fig_mol}
   \vspace{-20pt}
\end{figure}

According to \cite{moe}, we construct a DyNN with one gating network and $M$ parallel expert modules called mixture-of-experts. All expert modules have the same structure and are composed of two linear layers connected with the activation function. In details, the first linear layer has 32 neurons and uses the ReLU function. The second one has 30 neurons and uses the LogSoftmax function. The structure of DyNN is shown in Fig. \ref{fig_mol}. The task character is input to the gating network after processed by input layer. Then the gating network outputs the weight value to control each expert module. Therefore, when using DyNN for inference, only the output of the expert module with non-zero weight needs to be computed and multiplied by the corresponding weight. We denote by $Gate(x)_{i}$ and $E_{i}(x)$ the weight and output of the $i$-th expert module, respectively, and the output of DyNN can be described:
\begin{equation*}
    G(x;k)=\sum_{i=1}^{M} Gate(x;k)_{i} E_{i}(x) \tag{4}\label{Output}
\end{equation*}
Specifically, if the corresponding weight of an expert module is zero, the expert module will not perform computings. Through the operation of gating network, the expert modules can be screened, the more important ones will be strengthened, and the less effective ones will be weakened or removed. 

Obviously, when the weight of some expert modules is zero, the width and computation of DyNN will decrease, and thus the operation speed can be improved. Even when all weights are non-zero, the model performance is better than no-gating one due to the regulation of gating network on each expert module. If some of the expert modules are abandoned, the performance of the model will be affected due to the reduction of parameters. However, since the removed ones are those with least impacts on the overall model performance,it is likely to greatly reduce the computing delay with minor impact on the model accuracy. We have proved through experiments that removing some expert modules has little adverse impact on the accuracy of DyNN. The sparsity function is used as the gating network mapping, and the width of DyNN is determined by the given hyper-parameter $K$. Because the computational complexity of gating network is small, we ignore it for simplicity. A wide and deep gate network will have better performance when choosing expert network, but it will also increase the inference time of DyNN. In this paper we use a trainable parameter matrix $W_g$ to multiply the input $x$, and use the Softmax activation function at the end to keep the largest $K$ outputs as the selected expert modules, which can be represented by
\begin{equation*}
    Gate(x;K) = Softmax(TopK(W_gx,K)), \tag{5}
\end{equation*}
where $\operatorname{TopK}(v, k)_{i}$ is defined as 

\begin{equation*}
\operatorname{TopK}(v, k)_{i}=\left\{\begin{array}{ll}
v_{i} & \text { if } v_{i} \text { is in the top } k \text { elements of } v\tag{6} \\
-\infty & \text { otherwise. }
\end{array}\right.
\end{equation*}

\subsection{Two-Stage Network Optimization}
In the proposed mixture-of-experts DyNN, the width of the network is equivalent to the number of expert modules $K$, and therefore (3) can be transformed into
\begin{align*}
    & min_{\mathbf{b},\mathbf{\theta}}\quad T_{tra} + \sum_{i=1}^{N} w_i\frac{K \alpha \varphi}{f} \tag{7} \\
    &s.t.\quad\quad\quad \sum_{i} b_{i}\leq b_{max}, \tag{7a} \\
    &\quad\quad\quad\quad\quad \left\|\theta_{j}\right\| \textgreater 0\quad \forall j \in \{1,2,\cdots,l_e\}, \tag{7b}\\
    &\quad\quad\quad\quad\quad \mathbf{b} = G(\mathbf{w},\mathbf{s},\mathbf{d};\mathbf{\theta}), \tag{7c} \\
    &\quad\quad\quad\quad\quad \varphi=\sum_{j=1}^{l_e-1} \left\|\theta_{j}\right\| \left\|\theta_{j+1}\right\|, \tag{7d}
\end{align*}
where $\varphi$ is the computational complexity required to operate an expert module, and $l_e$ is the layers of one expert model. Due to constraint (3c), $T_{tra}$ and $T_{com}$ are a pair of coupled variables. We use a two-stage optimization method to solve this problem, which can be seen as an offline and online stage policy. In stage 1, we optimize the parameters of DyNN models with different $K$ values $(K\leq M)$, the DyNN model is trained. Then, we build a knowledge base about service delay and the network width to determine the best network width $K$.
\\
Stage-1:
We consider optimizing the parameters of DyNN network when $K$ is fixed, and then the optimization problem is described as
\begin{align*}
    & min_{\mathbf{b},\mathbf{\theta}}\quad T_{tra}\tag{8} \\ 
    &s.t.\quad\quad\quad \sum_{i} b_{i}\leq b_{max}, \tag{8a} \\
    &\quad\quad\quad\quad\quad \left\|\theta_{j}\right\| \textgreater 0\quad \forall j \in \{1,2,\cdots,K\}, \tag{8b}\\
    &\quad\quad\quad\quad\quad \mathbf{b} = G(\mathbf{w},\mathbf{s},\mathbf{d};\mathbf{\theta}), \tag{8c}
\end{align*}

In order to train the network to obtain the lowest transmission delay, we use the service transmission delay $T_{tra}$  as the loss function.

In order to ensure that the output of DyNN meets the constraint $\sum_{i} b_{i}=b_{\max }$, we use LogSoftmax as the activation function in the output layer. $K$ is randomly selected during training, and the network parameters are optimized by back propagation method.
\\
Stage-2:
In stage 1, we have completed the training of the DyNN model. Then, we build a knowledge base which show the optimal $K$-value under different task sizes and CPU frequencies. By comparing the network performance of all different $K$ values under a certain task size and CPU operation frequency, the optimal $K$ value in this case is obtained, and the optimal $K$ value knowledge base is constructed. By searching the knowledge base, the optimal $K$ can be found in a responsive manner, which deals with varying scenarios and cases, 
\begin{align*}
    & \min_{K}\quad T_{tra} + \sum_{i=1}^{N} w_i\frac{K \alpha \varphi}{f} \tag{9} \\
    &s.t.\quad\quad \varphi=\sum_{j=1}^{l_{e}-1} \left\|\theta_{j}\right\| \left\|\theta_{j+1}\right\|, \tag{9a}
\end{align*}

\subsection{Knowledge-Assisted DyNN}
After completing the optimization of DyNN parameters and the construction of the optimal $K$-value knowledge base, we adopt the knowledge-assisted DyNN scheme to realize the on-demand resource allocation. First, the scenario and service information, i.e., average task size and CPU frequency, is inputted into the knowledge base, and the optimal $K$ value is obtained. Then, we select the DyNN model with $K$ expert modules to conduct resource allocation inference, realizing the on-demand bandwidth allocation in the specific scenario.

\begin{algorithm}[!h]
	\caption{Two-Stage Optimization Algorithm}
	\begin{algorithmic}[1]
	\STATE Initialize $\mathbf{\theta}_0,r,epoch$, knowledge base of $K$

    \FOR{$i=0$ to epoch}
    \STATE Randomly select $K$ from \{1,2, \dots, M\} with equal probability
    \STATE $\mathbf{b}=G(\mathbf{s},\mathbf{w},\mathbf{d};\mathbf{\theta_{i}},K)$
    \STATE $Loss = \sum_{i}^{N} w_i s_i/(b_i \log_2 (1+\frac{P  (\frac{d_0}{d_{i}})^2 g_0}{\sigma^2} ))$
    \STATE $\theta_{i+1} = (1-r)\theta_i + r \arg\min_{\mathbf{\theta}} Loss$
    \ENDFOR
    \FOR{$s=s_{min}$ to $s_{max}$}
    \FOR{$f=f_{min}$ to $f_{max}$}
    \STATE $K = \arg\min_K T_{tra}+T_{com}$
    \STATE Record this $K$ value in the knowledge base $K$-Knowledge
    
    \ENDFOR
    \ENDFOR
	\end{algorithmic}
\end{algorithm}

\begin{algorithm}[!h]
	\caption{Knowledge Assisted DyNN Algorithm}
	\begin{algorithmic}[1]
	\STATE Input $s,w,f,d$
    \STATE Find the optimal $K$ value under $s,f$ in $K$-Knowledge
    \STATE $\mathbf{b}=G(\mathbf{s},\mathbf{w},\mathbf{d};\mathbf{\theta_i},K)$
    \STATE Return $\mathbf{b}$
	\end{algorithmic}
\end{algorithm}

\section{Simulation Results}

In this section, we evaluate the proposed DyNN-based resource allocation scheme using simulation. We set the simulation parameters as $g_{0}=-40 d B,\ d_{0}=1 m,\ P=0.1 W,\ \sigma^2=N_{0}b_{i}, \ N_{0}=10^{-17} W / H z,\ b_{max}=10 MHz$. In addition, we assume that the task size and transmission distance are uniformly distributed if not specified, i.e., $s_{i} \sim \text{U}\left(0, 10 Kb\right)$ and $d_{i } \sim \text{U}(0,500 m)$. For the importance of each task, we quantify it into four levels $w_{i} \in\{0.8, 0.4, 0.2, 0.1\}$ to measure different service urgency. In the following analysis, we use total weighted service delay $T_{tra} + T_{com}$ to measure the performance of the allocation algorithm. The maximum number of expert modules $M$ is set to 30. At the same time, in order to ensure the correctness of the comparison results, we use the same expert module to construct the static network, that is, the static network is identical to the DyNN except that it does not have the gating network. However, for other more diverse scenarios, how to choose the appropriate $M$ can also be a problem which still needs further research.

\subsection{Performance Evaluation of DyNN-based Scheme}

As explained in Section II, the reduction of network parameters will inevitably affect the model performance. To illustrate this, we compared the transmission delay of six DyNNs with different $K$ values and the static network in Table \ref{tab:comp} when the average data size is set to 10Kb. 
We observed that when the DyNN has the maximum width, its performance is better than the static network. This is due to the weighting effect of the gate network on the expert modules, making the training effect of the DyNN better under the same width. At the same time, it can also be found that with the decrease of $K$, the performance of the network does not significantly degrade. We define PERF in Table \ref{tab:comp} as
$\frac{\min T_{tra}}{T_{tra,i}}$ to simply measure the allocation performance of different models. It can be seen that even only one expert module is used, the DyNN can still achieve higher than 90 \% of the performance of the network with the maximum number of experts. Also, we give the amount of multiplication and accumulation (MACC) of different network models in the last row.

\begin{table}[h]
	\renewcommand{\arraystretch}{1.5}
	\newcolumntype{P}[1]{>{\centering\arraybackslash}p{#1}}
	\caption{The Relationship Between $T_{tra}$ and Algorithm Computing Amount on Network Width}
	\resizebox{\columnwidth}{!}{
	\begin{tabular}{c|ccccccc}
		\toprule[0.75pt]\toprule[0.75pt]
		$K$&	1&	2&	5&	10&	20&	30&	static\\\hline
		$T_{tra}(ms)$&	1.967&	1.922&	1.873&	1.859&	1.849&	1.841&	1.846\\\hline
		PERF&	93.59\%	&95.78\%&	98.29\%&	99.03\%	&99.57\%&	100\%&	99.73\%\\\hline
		MACC&	2.17K&	4.35K&	10.87K&	21.74K&	43.48k&	65.22K&	65.22K\\\bottomrule[0.75pt]\bottomrule[0.75pt]
	\end{tabular}}
	\label{tab:comp}
\end{table}

\begin{table}[h]
	\renewcommand{\arraystretch}{1.5}
	\newcolumntype{P}[1]{>{\centering\arraybackslash}p{#1}}
	\caption{$T_{tra}+T_{com}$ For Different Task Sizes in Different Width Networks}
	\resizebox{\columnwidth}{!}{
	\begin{tabular}{c|P{1cm}P{1cm}P{1cm}P{1cm}P{1cm}P{1cm}P{1cm}}
		\toprule[0.75pt]\toprule[0.75pt]
		\begin{normalsize}
		\diagbox[width=2cm,height=1cm,innerleftsep=0mm,innerrightsep=3mm]{Size(Kb)}{$K$}
		\end{normalsize}

	& 1     & 2     & 5     & 10    & 20    & 30    & \multicolumn{1}{c}{static} \\\hline
		2  & 0.392\quad & 0.389 & 0.393 & 0.408 & 0.439 & 0.470 & 0.471                      \\\hline
		4  & 0.773 & 0.761 & 0.757 & 0.769 & 0.798 & 0.829 & 0.831                      \\\hline
		6  & 1.159 & 1.14  & 1.128 & 1.137 & 1.163 & 1.194 & 1.195                      \\\hline
		8  & 1.571 & 1.539 & 1.517 & 1.524 & 1.549 & 1.574 & 1.577                      \\\hline
		10 & 1.971 & 1.928 & 1.889 & 1.892 & 1.914 & 1.939 & 1.943                \\\bottomrule[0.75pt]     \bottomrule[0.75pt]   
	\end{tabular}}
	\label{tab:comp-tra}
\end{table}

In Table \ref{tab:comp-tra}, by setting CPU frequency to 0.5GHz, we compare the performance of various DyNNs and static networks under different average task sizes. Obviously, the total weighted service delay increases linearly with task size. When the task is small, the required transmission delay is small, and therefore $T_{com}$ accounts for a large proportion of the total weighted delay. Therefore, the performance of DyNNs is significantly higher than that of the static network. Similarly, with the increase of task size, the proportion of $T_{tra}$ increases. In order to obtain lower delay, the allocation performance of the model (i.e., the transmission delay) is more important. Therefore, we can see that the advantages of DyNN are gradually decreasing. When the task size increases to 10Kb, the performance of static network has exceeded that of some DyNNs.

\begin{figure}[h]
  \centering
  \includegraphics[width=0.9\columnwidth]{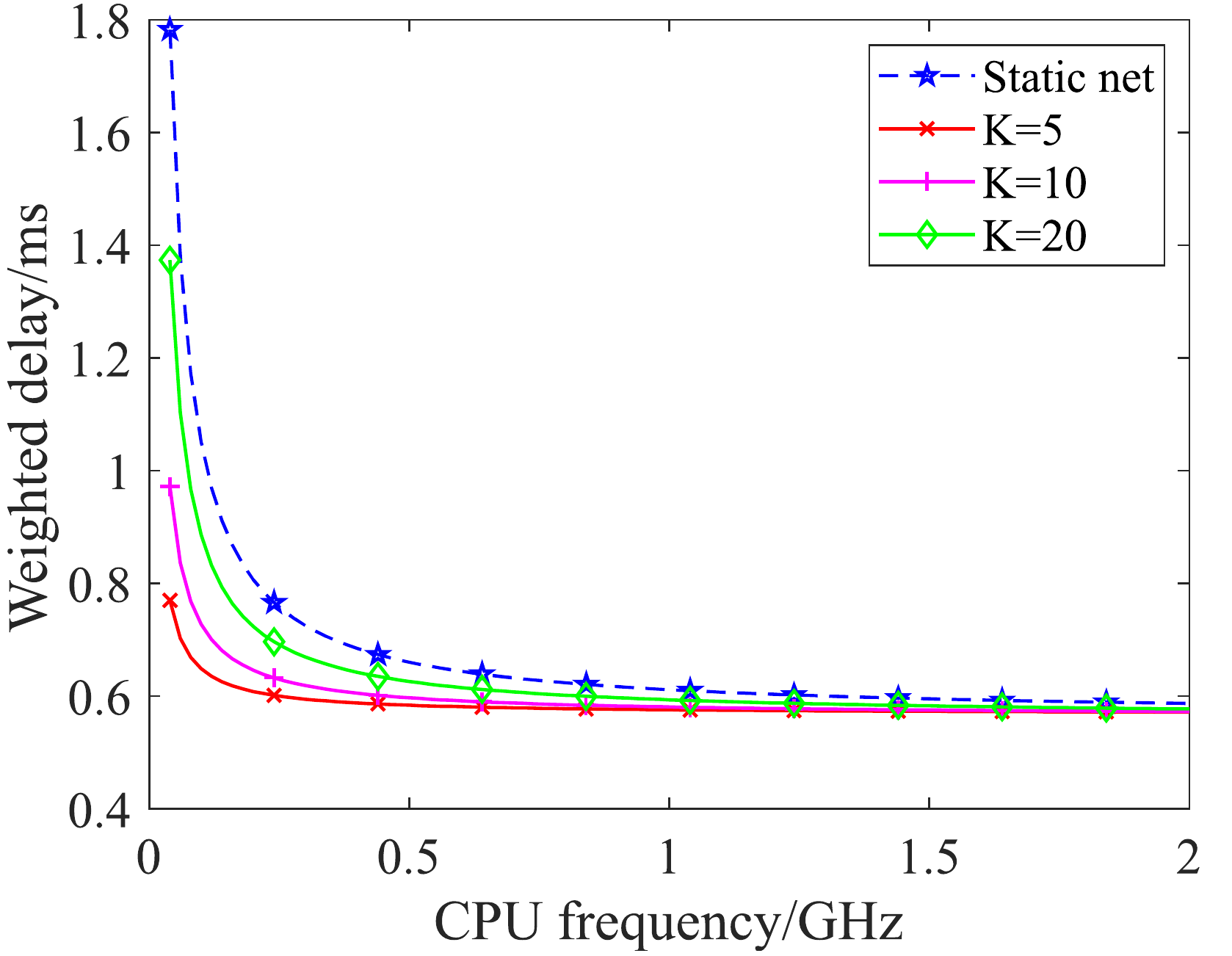}
   \vspace{-5pt}
  \caption {The relation of CPU frequency on weighted time.}
  \label{T-CPU}
   \vspace{-9pt}
\end{figure}

In Fig. \ref{T-CPU}, we show the relationship between $T_{tra}$ and base station CPU frequency. For comparison, we set the CPU frequency variation range to $\{20 M \sim 2 GHz\}$. It can be seen that $T_{tra}$ decreases with the increase of user CPU frequency. This is because the computing delay is inversely proportional to the user's computing capability. When the user's computing capability is weak, the computing delay accounts for a large proportion of the weight time. Therefore, the performance of the dynamic network with small amount of computing is significantly better than that of the static network, especially when $K$ is small. With the increase of computing capability, the inference delay decreases, and the total weight delay depends more on the transmission delay. Therefore, the advantage of dynamic network is weakened. Theoretically, when the base station has enough computing capability such that the computing delay can be ignored, the dynamic network with large $K$ or static network will achieve the optimal performance.

\begin{figure}[h]
  \centering
  \includegraphics[width=0.9\columnwidth]{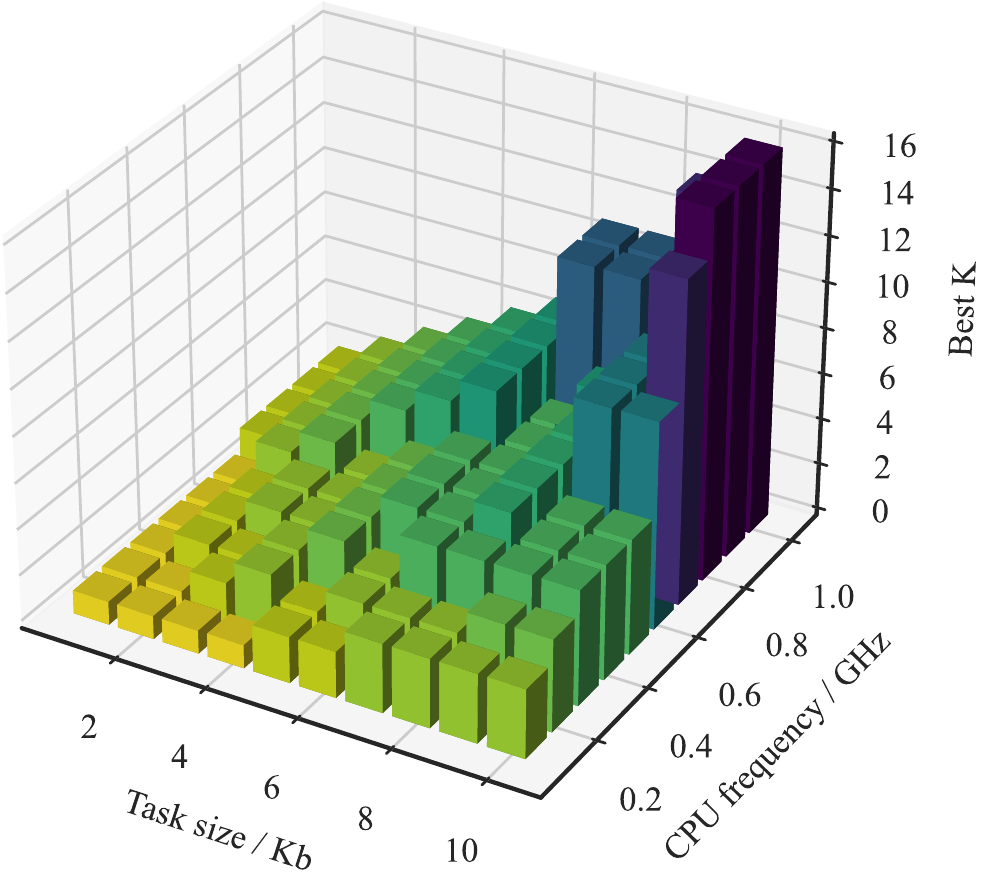}
   \vspace{-5pt}
  \caption {The best $K$ value under different CPU frequency and task size.}
  \label{knowledge}
   \vspace{-13pt}
\end{figure}

\subsection{Further Optimization Based on Knowledge-Assisted DyNN}
In the proposed DyNN-based scheme, the selection of the value of $K$ greatly impacts the overall performance. In order to clearly show the selection of DyNN's width under different scenarios, we draw the three-dimensional figure of the optimal $K$ value with CPU frequency and task size in Fig. \ref{knowledge}. It can be seen that in the case of low frequency and small tasks, it is a good choice to use a narrow network. With the increase of task size and the improvement of computing capability, it is wiser to select a wider network to achieve higher performance. It is the ability of selecting the proper width based on the specific requirements and scenario which gives the DyNN advantages. To fully exploit the advantages of DyNN, we construct a knowledge base using which the optimal $K$ can be efficiently determined.

\begin{figure}[h]
  \centering
  \includegraphics[width=0.9\columnwidth]{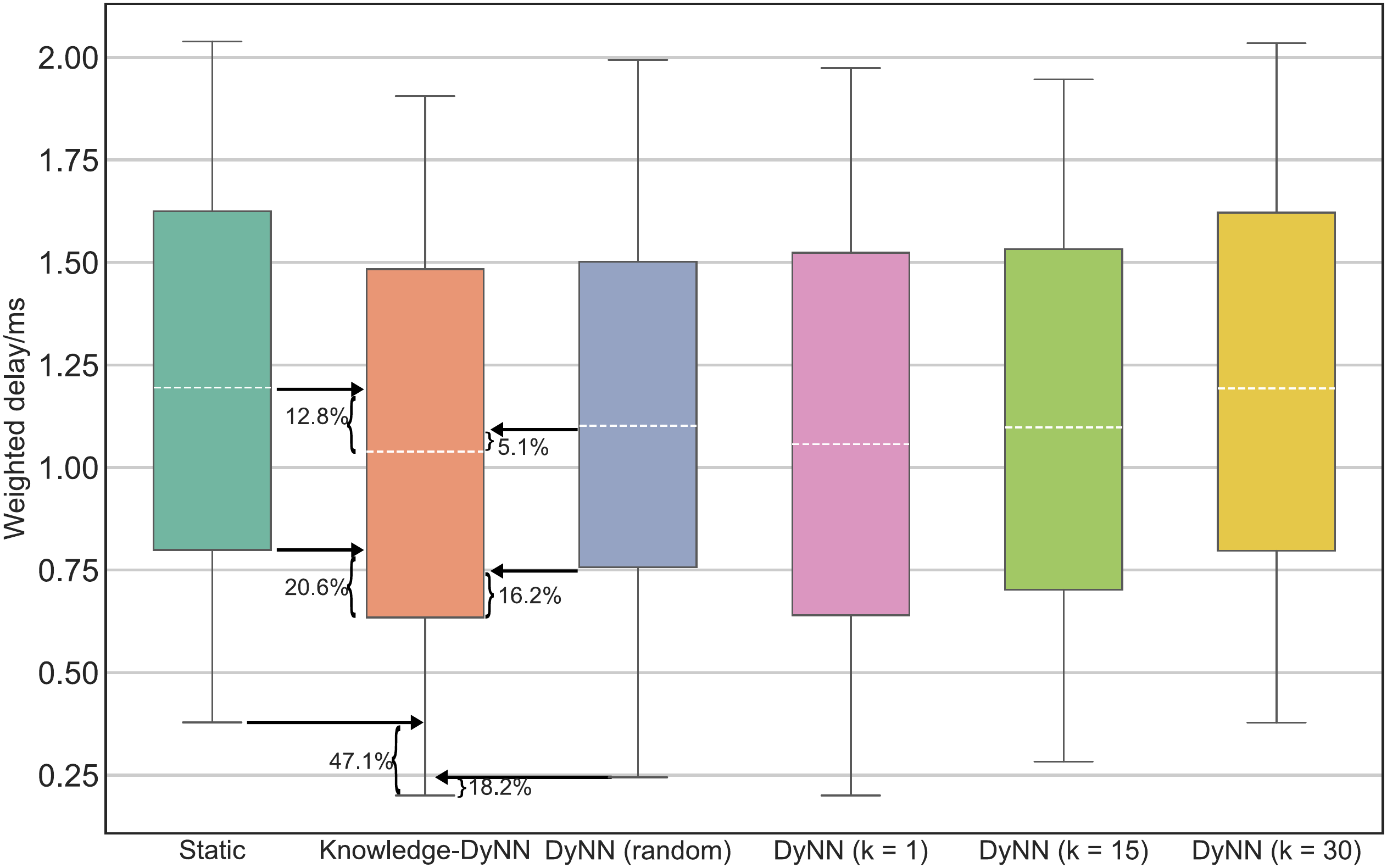}
   \vspace{-5pt}
  \caption {The performance of different algorithm under random CPU frequency and task size.}
  \label{boxplot}
   \vspace{-12pt}
\end{figure}

To demonstrate the advantages of knowledge, Fig. \ref{boxplot} shows the performance comparison of static network, fixed $K$-value DyNN and knowledge-assisted DyNN under different scenarios.
Similar to Fig. \ref{knowledge}, we set the task size to $\{0 \sim 10Kb\}$ at equal intervals. The CPU frequency of each user is randomly selected from $\{10 M \sim 1 GHz\}$. It can be seen that the performance of knowledge-assisted DyNN achieves much higher performance than the static network and fixed $K$-value DyNN. Compared with the static network, the minimum delay of Knowledge DyNN is reduced by $47.1\%$, reflecting the advantages in the case of small tasks. Even when considering large tasks, we can still observe obvious performance improvement, reducing the delay of median and lower quantile by $12.8\%$ and $20.6\%$, respectively. Knowledge DyNN is always better than fixed or random $K$-value DyNN, reflecting the advantages for knowledge.

\section{Conclusion}
In this paper, we have investigated the on-demand resource orchestration problem in multi-service wireless networks. Using knowledge-based dynamic neural networks, we have proposed a flexible bandwidth allocation algorithm with adjustive computational complexity and optimality performance. Simulation results have shown that the proposed scheme achieves better performance than traditional static neural network-based methods, especially when the task is small. Due to the dynamic selection of network width in different deployment situations, the model can greatly balance inference delay and allocation performance, so as to obtain improved flexibility. By applying the proposed scheme in the network, the network services are capable of adjusting the service delay and resource orchestration performance according to the service requirements and computing power. For futher research, we will explore more novel dynamic neural networks, and apply the DyNN-based methods in more sophisticated network resource orchestration and management problems.

\section*{Acknowledgement}
This work was supported in part by the National Key R$\&$D Program of China under Grant 2020YFB1807700, in part by the National Natural Science Foundation of China (NSFC) under Grant 62071356, and in part by the Fundamental Research Funds for the Central Universities under Grant JB210113.

\bibliography{ref}
\bibliographystyle{IEEEtran}

\ifCLASSOPTIONcaptionsoff
  \newpage
\fi

\end{document}